\documentclass[copyright,creativecommons,hidelinks]{eptcs}
\def\event{EXPRESS/SOS 2020}
\usepackage{latexsym,underscore,amsmath,amssymb,color,wrapfig}
\newfont{\bbb}{bbm10 scaled 1100}                       
\newcommand{\IN}{\mbox{\bbb N}}                         
\newcommand{\IP}{\mbox{\bbb P}}                         
\newcommand{\IT}{\mbox{\bbb T}}                         
\DeclareSymbolFont{frenchscript}{OMS}{ztmcm}{m}{n}
\DeclareMathSymbol{\Pow}{\mathord}{frenchscript}{80}    
\DeclareMathSymbol{\A}{\mathord}{frenchscript}{65}      
\DeclareMathSymbol{\Ce}{\mathord}{frenchscript}{67}     
\DeclareMathSymbol{\K}{\mathord}{frenchscript}{75}      
\DeclareMathSymbol{\Tsk}{\mathord}{frenchscript}{84}    
\renewcommand{\phi}{\varphi}
\newcommand{\dcup}{\stackrel{\mbox{\huge .}}{\cup}}   
\makeatletter
\newif\if@qeded
\global\@qededtrue
\def\qed{\hfill$\Box$\global\@qededtrue}
\def\qedneeded{\global\@qededfalse}
\def\qedifneeded{\if@qeded\else\qed\fi}
\makeatother
\newtheorem{defi}{Definition}
\newtheorem{theo}{Theorem}
\newtheorem{prop}{Proposition}
\newtheorem{lemm}{Lemma}
\newtheorem{coro}{Corollary}
\newtheorem{exam}{Example}
\newtheorem{obse}{Observation}

\newenvironment{definition}[1]{\begin{defi} \rm \label{df:#1} }{\end{defi}}
\newenvironment{definitionq}[2]{\begin{defi}[#2] \rm \label{df:#1} }{\end{defi}}

\newenvironment{example}[1]{\begin{exam} \rm \label{ex:#1} }{\end{exam}}

\newcommand{\df}[1]{Definition~\ref{df:#1}}

\newcommand{\ex}[1]{Example~\ref{ex:#1}}

\newcommand{\Sec}[1]{Section~\ref{sec:#1}}
\makeatletter
\def\comesfrom{\@transition\leftarrowfill}
\def\goesto{\@transition\rightarrowfill}
\def\ngoesto{\@transition\nrightarrowfill}
\def\Goesto{\@transition\Rightarrowfill}
\def\nGoesto{\@transition\nRightarrowfill}
\def\xmapsto{\@transition\mapstofill}
\def\nxmapsto{\@transition\nmapstofill}
\def\@transition#1{\@@transition{#1}}
\newbox\@transbox
\newbox\@arrowbox
\def\@@transition#1#2%
   {\setbox\@transbox\hbox
      {\vrule height 1.5ex depth .8ex width 0ex\hskip0.25em$\scriptstyle#2$\hskip0.25em}
   \ifdim\wd\@transbox<1.5em
      \setbox\@transbox\hbox to 1.5em{\hfil\box\@transbox\hfil}\fi
   \setbox\@arrowbox\hbox to \wd\@transbox{#1}
   \ht\@arrowbox\z@\dp\@arrowbox\z@
   \setbox\@transbox\hbox{$\mathop{\box\@arrowbox}\limits^{\box\@transbox}$}
   \dp\@transbox\z@\ht\@transbox 12pt
   \mathrel{\box\@transbox}}
\def\nrightarrowfill{$\m@th\mathord-\mkern-6mu%
  \cleaders\hbox{$\mkern-2mu\mathord-\mkern-2mu$}\hfill
  \mkern-6mu\mathord\not\mkern-2mu\mathord\rightarrow$}
\def\Rightarrowfill{$\m@th\mathord=\mkern-6mu%
  \cleaders\hbox{$\mkern-2mu\mathord=\mkern-2mu$}\hfill
  \mkern-6mu\mathord\Rightarrow$}
\def\nRightarrowfill{$\m@th\mathord=\mkern-6mu%
  \cleaders\hbox{$\mkern-2mu\mathord=\mkern-2mu$}\hfill
  \mkern-6mu\mathord\not\mathord\Rightarrow$}
\def\mapstofill{$\m@th\mathord\mapstochar\mathord-\mkern-6mu%
  \cleaders\hbox{$\mkern-2mu\mathord-\mkern-2mu$}\hfill
  \mkern-6mu\mathord\rightarrow$}
\def\nmapstofill{$\m@th\mathord\mapstochar\mathord-\mkern-6mu%
  \cleaders\hbox{$\mkern-2mu\mathord-\mkern-2mu$}\hfill
  \mkern-6mu\mathord\not\mkern-2mu\mathord\rightarrow$}
\makeatother 
\newcommand{\goto}[2][]{\mathrel{\goesto{~#2{\color{red}\;,\;#1}~}}}        
\newcommand{\weg}[1]{}                                
\newcommand{\plat}[1]{\raisebox{0pt}[0pt][0pt]{#1}}   
\newcommand{\LTLX}{LTL$^{}_{\!\!\bf -X}$}           
\newcommand{\Tr}{\textit{Tr}}                         
\newcommand{\source}{\textit{source\/}}               
\newcommand{\target}{\textit{target\/}}               
\def\powermultiset#1{\IN^{#1}}
\newcommand{\monus}{\mathrel{\raisebox{-0pt}[0pt][0pt]{$
                      \stackrel{\raisebox{-5pt}[0pt][0pt]{\huge$\cdot$}}
                               {\raisebox{0pt}[0pt][0pt]{$-$}}$}}}
\def\precond#1{{\vphantom{#1}}^\bullet #1}
\def\postcond#1{{#1}^\bullet}
\newcommand{\cT}{{\rm T}}                             
\newcommand{\Left}{\textsc{l}}                        
\newcommand{\R}{\textsc{r}}                           
\newcommand{\nconc}{\,\not\!\smile}                   

\newcommand{\lni}[1][i]{\mbox{\color{blue}\it ln$_{#1}$}}
\newcommand{\en}[1][i]{\mbox{\color{blue}\it en$_{#1}$}}
\newcommand{\lc}[1][i]{\mbox{\color{red}\it lc$_{#1}$}}
\newcommand{\ec}[1][i]{\mbox{\color{red}\it ec$_{#1}$}}

\def\titlerunning{Reactive Temporal Logic}
\title{\titlerunning}
\author{Rob van Glabbeek
\institute{Data61, CSIRO, Sydney, Australia}
\institute{School of Computer Science and Engineering,
University of New South Wales, Sydney, Australia}
\email{rvg@cs.stanford.edu}
}
\def\authorrunning{R.J. van Glabbeek}
\begin{document}
\maketitle

\begin{abstract}
Whereas standard treatments of temporal logic are adequate for \emph{closed systems}, having no
run-time interactions with their environment, they fall short for \emph{reactive systems},
interacting with their environments through synchronisation of actions.
This paper introduces \emph{reactive temporal logic}, a form of temporal logic adapted
for the study of reactive systems. I illustrate its use by applying it to formulate definitions
of a fair scheduler, and of a correct mutual exclusion protocol. Previous definitions of these
concepts were conceptually much more involved or less precise, leading to debates on whether or  not
a given protocol satisfies the implicit requirements.
\end{abstract}

\section{Introduction}

\emph{Labelled transition systems} are a common model of distributed systems.  They consist of sets
of states, also called \emph{processes}, and transitions---each transition going from a source state
to a target state. A given distributed system $\mathcal{D}$ corresponds to a state $P$ in a transition system
$\IT$---the initial state of $\mathcal{D}$.  The other states of $\mathcal{D}$ are the processes in $\IT$
that are reachable from $P$ by following the transitions. The transitions are labelled by
\emph{actions}, either visible ones or the invisible action $\tau$. Whereas a $\tau$-labelled
transition represents a state-change that can be made spontaneously by the represented system,
$a$-labelled transitions for $a\neq \tau$ merely represent potential activities of $\mathcal{D}$,
for they require cooperation from the \emph{environment} in which $\mathcal{D}$ will be running, sometimes
identified with the \emph{user} of system $\mathcal{D}$. A typical example is the acceptance of a
coin by a vending machine. For this transition to occur, the vending machine should be in a state
where it is enabled, i.e., the opening for inserting coins should not be closed off,
but also the user of the system should partake by inserting the coin.

{\makeatletter
\let\par\@@par
\par\parshape0
\everypar{}\begin{wrapfigure}[3]{r}{0.2\textwidth}
 \vspace{-4ex}
 \input{pretzel}
  \centerline{\raisebox{1ex}{\box\graph}}
 \end{wrapfigure}
Consider a vending machine that alternatingly accepts a coin ($c$) and produces a pretzel ($p$).
Its labelled transition system is depicted on the right. In standard temporal logic one can express
that each action $c$ is followed by $p$: whenever a coin is inserted, a pretzel will be produced.
Aligned with intuition, this formula is valid for the depicted system.
However, by symmetry one obtains the validity of a formula saying that each $p$ is followed by a
$c$: whenever a pretzel is produced, eventually a new coin will be inserted. But that clashes with intuition.
\par}
In this paper I enrich temporal logic judgements $P \models \phi$, saying that system $P$ satisfies
formula $\phi$, with a third argument $B$, telling which actions can be blocked by the environment
(by failing to act as a synchronisation partner) and which cannot. When stipulating that the coin
needs cooperation from a user, but producing the pretzel does not, the two temporal judgements can be
distinguished, and only one of them holds. I also introduce a fourth argument $CC$---a completeness
criterion---that incorporates progress, justness and fairness assumptions employed when making a
temporal judgement. This yields statements of the form $P \models^{CC}_B \phi$.

Then I use the so obtained formalism to formalise the correctness requirements of mutual exclusion
protocols and of fair schedulers. Making these requirements precise helps in stating negative
results on the possibilities to render such protocols in a given setting. In the case of fair
schedulers, reactive temporal logic leads to a much easier to understand formalisation than the one
in the literature. In the case of mutual exclusion protocols it leads to more precise and less
ambiguous requirements, that may help to settle debates on whether or not some formalisation of a
mutual exclusion protocol is correct.

\section{Kripke Structures and Linear-time Temporal Logic}\label{sec:Kripke}

\begin{definition}{Kripke}
Let $AP$ be a set of \emph{atomic predicates}.
A \emph{Kripke structure} over $AP$ is tuple $(S, \rightarrow, \models)$ with $S$ a set (of \emph{states}),
${\rightarrow} \subseteq S \times S$, the \emph{transition relation}, and ${\models} \subseteq S \times AP$.
$s \models p$ says that predicate $p\in AP$ \emph{holds} in state $s \in S$.
\end{definition}
Here I generalise the standard definition \cite{HuthRyan04} by dropping the condition of \emph{totality},
requiring that for each state $s\in S$ there is a transition $(s,s')\in {\rightarrow}$.
A \emph{path} in a Kripke structure is a nonempty finite or infinite sequence $s_0,s_1,\dots$ of states, such
that $(s_i,s_{i+1}) \in {\rightarrow}$ for each adjacent pair of states $s_i,s_{i+1}$ in that sequence.
A \emph{suffix} $\pi'$ of a path $\pi$ is any path obtained from $\pi$ by removing an initial segment.
Write $\pi \Rightarrow \pi'$ if $\pi'$ is a suffix of $\pi$; this relation is reflexive and transitive.

A distributed system $\mathcal{D}$ can be modelled as a state $s$ in a Kripke structure $K$.
A run of $\mathcal{D}$ then corresponds with a path in $K$ starting in $s$.
Whereas each finite path in $K$ starting from $s$ models a \emph{partial run} of $\mathcal{D}$,
i.e., an initial segment of a (complete) run, typically not each path models a run.
Therefore a Kripke structure constitutes a good model of distributed systems
only in combination with a \emph{completeness criterion} \cite{vG19}: a selection of a
set of paths as \emph{complete paths}, modelling runs of the represented system.

The default completeness criterion, implicitly used in almost all work on temporal logic, classifies
a path as complete iff it is infinite. In other words, only the infinite paths, and all of them,
model (complete) runs of the represented system. This applies when adopting the condition of
totality, so that each finite path is a prefix of an infinite path.  Naturally, in this setting
there is no reason to use the word ``complete'', as ``infinite'' will do.  As I plan to discuss
alternative completeness criteria in \Sec{completeness criteria}, I will here already refer to paths
satisfying a completeness criterion as ``complete'' rather than ``infinite''.
Moreover, when dropping totality, the default completeness criterion is adapted to declare a path
complete iff it either is infinite or ends in a state without outgoing transitions \cite{DV95}.

\emph{Linear-time temporal logic} (LTL) \cite{Pnueli77,HuthRyan04} is a formalism explicitly designed to formulate
properties such as the safety and liveness requirements of mutual exclusion protocols. Its syntax is
\[\phi,\psi ::= p \mid \neg \phi \mid \phi \wedge \psi  \mid {\bf X}\phi \mid  {\bf F}\phi \mid
           {\bf G}\phi \mid  \psi {\bf U} \phi \]
with $p \in AP$ an atomic predicate. The propositional connectives $\Rightarrow$ and $\vee$ can be
added as syntactic sugar. It is interpreted on the paths in a Kripke structure.
The relation $\models$ between paths and LTL formulae, with $\pi\models \phi$ saying that the path
$\pi$ \emph{satisfies} the formula $\phi$, or that $\phi$ is \emph{valid} on $\pi$, is inductively
defined by
\begin{itemize}
\item $\pi \models p$, with $p\in AP$, iff $s\models p$, where $s$ is the first state of $\pi$,
\item $\pi \models \neg\phi$ iff $\pi \not\models \phi$, 
\item $\pi \models \phi \wedge \psi$ iff $\pi \models \phi$ and $\pi \models \psi$, 
\item $\pi \models {\bf X} \phi$ iff $\pi'\models\phi$, where $\pi'$ is the suffix of $\pi$ obtained by
  omitting the first state,
\item $\pi \models {\bf F} \phi$ iff $\pi'\models\phi$ for some suffix $\pi'$ of $\pi$,
\item $\pi \models {\bf G} \phi$ iff $\pi'\models\phi$ for each suffix $\pi'$ of $\pi$, and
\item $\pi \models \psi {\bf U} \phi$ iff $\pi'\models\phi$ for some suffix $\pi'$ of $\pi$,
  and $\pi'' \models \psi$ for each path $\pi''\neq \pi'$ with $\pi\Rightarrow\pi''\Rightarrow\pi'$.
\end{itemize}
In \cite{Lam83}, Lamport argues against the use of the next-state operator {\bf X}, as it is incompatible
with abstraction from irrelevant details in system descriptions. Following this advice,
I here restrict attention to LTL without the next-state modality, {\LTLX}.

In the standard treatment of LTL \cite{Pnueli77,HuthRyan04}, judgements $\pi\models \phi$ are
pronounced only for infinite paths $\pi$. Here I apply the same definitions verbatim to
finite paths as well. At this point I benefit from the exclusion of the next-state operator {\bf X}.
In its presence I would have to decide what is the meaning of a judgement $\pi \models {\bf X}\phi$
when $\pi$ is a path consisting of a single state.\footnote{One possibility would be to declare this judgement
to be false, regardless of $\phi$. However, this would invalidate the self-duality of the {\bf X}
modality, stating that $\neg{\bf X}\phi$ holds for the same paths as ${\bf X}\neg\phi$.}

Having given meaning to judgements $\pi \models \phi$, as a derived concept one defines when an
{\LTLX} formula $\phi$ holds for a state $s$ in a Kripke structure, modelling a distributed system
$\mathcal{D}$, notation $s \models \phi$ or $\mathcal{D} \models \phi$. This is the case iff $\phi$
holds for all runs of $\mathcal{D}$.
\begin{definition}{validity}
$s \models \phi$ iff $\pi \models \phi$ for all complete paths $\pi$ starting in state $s$.
\end{definition}
Note that this definition depends on the underlying completeness criterion, telling which paths
model actual system runs. In situations where I consider different completeness criteria, I make this
explicit by writing $s \models^{CC} \phi$, with $CC$ the name of the completeness criterion used.
When leaving out the superscript $CC$ I here refer to the default completeness criterion, defined above.

\begin{example}{beer}
Alice, Bart and Cameron stand behind a bar, continuously ordering and drinking beer.
Assume they do not know each other and order individually.
As there is only one barman, they are served sequentially.
Also assume that none of them is served twice in a row, but as it takes no longer to drink a beer
than to pour it, each of them is ready for the next beer as soon as another person is served.
{\makeatletter
\let\par\@@par
\par\parshape0
\everypar{}\begin{wrapfigure}[6]{r}{0.25\textwidth}
 \vspace{-1ex}
 \input{Bart}
  \centerline{\raisebox{1ex}{\box\graph}}
 \end{wrapfigure}

A Kripke structure of this distributed system $\mathcal{D}$ is drawn on the right.
The initial state of $\mathcal{D}$ is indicated by a short arrow. The other three states are
labelled with the atomic predicates $A$, $B$ and $C$, indicating that Alice, Bart or Cameron,
respectively, has just acquired a beer. When assuming the default completeness criterion, valid
{\LTLX} formulae are ${\bf F}(A \vee C)$, saying that eventually either Alice or Cameron will get a
beer, or ${\bf G}(A \Rightarrow {\bf F}\neg A)$, saying that each time Alice got a beer
is followed eventually by someone else getting one. However, it is not guaranteed that Bart will
ever get a beer: $\mathcal{D} \not\models {\bf F}B$. A counterexample for this formula is the infinite run in
which Alice and Cameron get a beer alternatingly.
\par}
\end{example}

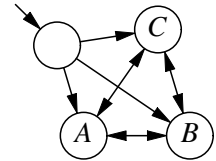
\begin{example}{Bart alone}
Bart is the only customer in a bar in London, with a single barman.
He only wants one beer.
{\makeatletter
\let\par\@@par
\par\parshape0
\everypar{}\begin{wrapfigure}[6]{r}{0.25\textwidth}
 \vspace{-1.5ex}
 \input{Bart2}
  \centerline{\raisebox{1ex}{\box\graph}}
 \end{wrapfigure}
\noindent
A Kripke structure of this system $\mathcal{E}$ is drawn on the right.
When assuming the default completeness criterion, this time Bart gets his beer:
$\mathcal{E} \models {\bf F}B$.
\par}
\end{example}

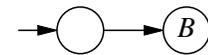
\begin{example}{Bart separated}
Bart is the only customer in a bar in London, with a single barman.
He only wants one beer.
{\makeatletter
\let\par\@@par
\par\parshape0
\everypar{}\begin{wrapfigure}[4]{r}{0.25\textwidth}
 \vspace{-1ex}
 \input{Bart3}
  \centerline{\raisebox{1ex}{\box\graph}}
 \end{wrapfigure}
\noindent
At the same time, Alice and Cameron are in a bar in Tokyo.
They drink a lot of beer. Bart is not in contact with Alice and Cameron,
nor is there any connection between the two bars.
Yet, one may choose to model the drinking in these two bars as a single distributed system.
A Kripke structure of this system $\mathcal{F}$ is drawn on the right, collapsing the orders of
Alice and Cameron, which can occur before or after Bart gets a beer, into self-loops.
When assuming the default completeness criterion, Bart cannot count on a beer:
$\mathcal{F}\not\models {\bf F}B$.
\par}
\end{example}

\section{Labelled Transition Systems, Process Algebra and Petri Nets}\label{sec:Models}

The most common formalisms in which to present reactive distributed systems are pseudocode,
process algebra and Petri nets. The semantics of these formalisms is often given by translation into
labelled transition systems (LTSs), and these in turn can be translated into Kripke structures, on
which temporal formulae from languages such as LTL are interpreted. These translations make the
validity relation $\models$ for temporal formulae applicable to all these formalisms. A state in
an LTS, for example, is defined to satisfy an {\LTLX} formula $\phi$ iff its translation into a
state in a Kripke structure satisfies this formula.

\begin{figure}[ht]
\input{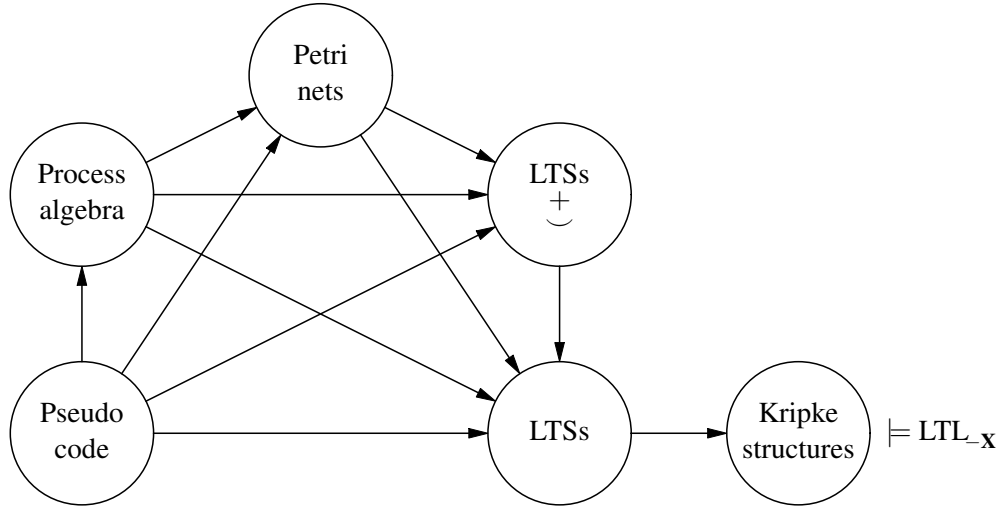}
\centerline{\raisebox{1ex}{\box\graph}}
\caption{\it Formalisms for modelling mutual exclusion protocols}\label{models}
\end{figure}

\noindent
Figure~\ref{models} shows a commuting diagram of semantic translations found in the literature, from
pseudocode, process algebra and Petri nets via LTSs to Kripke structures. Each step in the
translation abstracts from certain features of the formalism at its source.
Some useful requirements on distributed systems can be adequately formalised in
process algebra or Petri nets, and informally described for pseudocode, whereas LTSs and Kripke
structures have already abstracted from the relevant information. An example will be FS\hspace{1pt}1 on page \pageref{FS1}.
I also consider LTSs upgraded with a concurrency relation $\smile$ between transitions; these will be
expressive enough to formalise some of these requirements.

\subsection{Labelled Transition Systems}

\begin{definition}{LTS}
Let $A$ be a set of \emph{observable actions}, and let $Act := A \cup\{\tau\}$, with $\tau\notin A$
the \emph{hidden action}.
A \emph{labelled transition system} (LTS) over $Act$ is tuple $(\IP,  \Tr, \source,\target,\ell)$
with $\IP$ a set (of \emph{states} or \emph{processes}),
$\Tr$ a set (of \emph{transitions}), $\source,\target:\Tr\rightarrow \IP$ and $\ell:\Tr\rightarrow Act$.
\end{definition}
Write $s \goesto{\alpha} s'$ if there exists a transition $t$ with $\source(t)=s \in\IP$,
$\ell(t)=\alpha\in Act$ and $\target(t)=s'\in\IP$. In this case $t$ \emph{goes} from $s$ to $s'$,
and is an \emph{outgoing transition} of $s$.
A \emph{path} in an LTS is a finite or infinite alternating sequence of states and transitions,
starting with a state, such that each transition goes from the state before it to the state after it
(if any). A \emph{completeness criterion} on an LTS is a set of its paths.

As for Kripke structures, a distributed system $\mathcal{D}$ can be modelled as a state $s$ in an
LTS upgraded with a completeness criterion. A (complete) run of $\mathcal{D}$ is then modelled 
by a complete path starting in $s$. As for Kripke structures, the default completeness criterion
deems a path complete iff it either is infinite or ends in a \emph{deadlock}, a state without
outgoing transitions. An alternative completeness criterion could declare some infinite paths
incomplete, saying that they do not model runs that can actually occur, and/or declare some
finite paths that do not end in deadlock complete. A complete path $\pi$ ending in a state
models a run of the represented system that follows the path until its last state, and then stays
in that state forever, without taking any of its outgoing transitions. A complete path that ends in
a transition models a run in which the action represented by this last transition starts occurring but
never finishes. It is often assumed that transitions are instantaneous, or at least of finite duration.
This assumption is formalised through the adoption of a completeness criterion that holds all paths
ending in a transition to be incomplete.

The most prominent translation from LTSs to Kripke structures is from De Nicola \& Vaandrager \cite{DV95}.
Its purpose is merely to efficiently lift the validity relation $\models$ from Kripke structures to LTSs.
It simply creates a new state halfway along any transition labelled by a visible action, and moves the
transition label to that state.
\begin{definition}{DV translation}
Let $(\IP,  \Tr, \source,\target,\ell)$ be an LTS over $Act = A \cup\{\tau\}$. The associated Kripke structure
$(S,\rightarrow,\models)$ over $A$ is given by
\begin{itemize}
\item $S := \IP \cup \{t \in\Tr\mid \ell(t)\neq\tau\}$,
\item ${\rightarrow} := \{(source(t),t),(t,target(t)) \mid t \in\Tr\wedge \ell(t)\neq\tau\} \cup 
      \{(\source(t),\target(t))\mid t\in\Tr \wedge \ell(t)=\tau\}$
\item and ${\models} := \{(t,\ell(t)) \mid t \in \Tr \wedge \ell(t)\neq\tau\}$.
\end{itemize}
\end{definition}
Ignoring paths ending within a $\tau$-transition, which are never deemed complete anyway,
this translation yields a bijective correspondence between the paths in an LTS and the path in its
associated Kripke structure. Consequently, any completeness criterion on the LTS induces a
completeness criterion on the Kripke structure.
Hence it is now well-defined when $s \models^{CC} \phi$, with $s$ a state in an LTS, $CC$ a completeness
criterion on this LTS and $\phi$ an {\LTLX} formula.

\subsection{Petri Nets}\label{sec:nets}

\begin{definition}{net}
  A \emph{(labelled) Petri net} over $Act$ is a tuple
  $N = (S, T, F, M_0, \ell)$ where
  \begin{itemize}
    \item $S$ and $T$ are disjoint sets (of \emph{places} and \emph{transitions}),
    \item $F: (S \mathord\times T \mathrel\cup T \mathord\times S) \rightarrow \IN$
      (the \emph{flow relation} including \emph{arc weights})
      such that  $\forall t \mathbin\in T  \exists s\mathbin\in S.~ F(s, t) > 0$,
    \item $M_0 : S \rightarrow \IN$ (the \emph{initial marking}), and
    \item $\ell: T \rightarrow Act$ (the \emph{labelling function}).
  \end{itemize}
\end{definition}

\noindent
Petri nets are depicted by drawing the places as circles and the transitions as boxes, containing
their label. For $x,y \mathbin\in S\cup T$ there are $F(x,y)$ arrows (\emph{arcs}) from $x$ to $y$.
When a Petri net represents a distributed system, a global state of this system is given as a
\emph{marking}, a multiset of places, depicted by placing $M(s)$ dots (\emph{tokens}) in each place
$s$.  The initial state is $M_0$.  The behaviour of a Petri net is defined by the possible moves between
markings $M$ and $M'$, which take place when a finite multiset $G$ of transitions \emph{fires}.
In that case, each occurrence of a transition $t$ in $G$ consumes $F(s,t)$ tokens from each place $s$.
Naturally, this can happen only if $M$ makes all these tokens available in the first place. Next,
each $t$ produces $F(t,s)$ tokens in each $s$. \df{firing} formalises this notion of behaviour.

A {\em multiset} over a set $X$ is a function $A\!:X \rightarrow \IN$, i.e.\ $A\in \powermultiset{X}\!\!$.
Object $x \in X$ is an \emph{element of} $A$ iff $A(x) > 0$.
A multiset is \emph{empty} iff it has no elements, and \emph{finite} iff the set of its elements in finite.
For multisets $A$ and $B$ over $X$ I write $A \leq B$ iff \mbox{$A(x) \leq B(x)$} for all $x \mathbin\in X$;
$A + B$ denotes the multiset over $X$ with $(A + B)(x):=A(x)+B(x)$,
$A - B$ is given by
$(A - B)(x):=A(x)\monus B(x)=\mbox{max}(A(x)-B(x),0)$, and
for $k\mathbin\in\IN$ the multiset $k\cdot A$ is given by
$(k \cdot A)(x):=k\cdot A(x)$.
With $\{x,x,y\}$ I denote a multiset $A$ with
$A(x)\mathbin=2$ and $A(y)\mathbin=1$, rather than the set $\{x,y\}$ itself.

\begin{definition}{preset}
  Let $N\!=\!(S, T, F, M_0, \ell)$ be a Petri net and $t\in T$.
The multisets $\precond{t},~\postcond{t}: S \rightarrow \IN$ are given by $\precond{t}(s)=F(s,t)$ and
$\postcond{t}(s)=F(t,s)$ for all $s \in S$.
The elements of $\precond{t}$ and $\postcond{t}$ are
called \emph{pre-} and \emph{postplaces} of $t$, respectively.
These functions extend to finite multisets
$G: T \rightarrow\IN$ as usual, by
$\precond{G} := \sum_{t \in T}G(t)\cdot\precond{t}$ and
$\postcond{G} := \sum_{t \in T}G(t)\cdot\postcond{t}$.
\end{definition}

\begin{definition}{firing}
  Let $N \mathbin= (S, T, F, M_0,\ell)$ be a Petri net,
  $G \in \IN^T\!$, $G$ non-empty and finite, and $M, M' \in \IN^S\!$.
\\
$G$ is a \emph{step} from $M$ to $M'$,
written $M\goesto{G}_N M'$, iff
$^\bullet G \leq M$ ($G$ is \emph{enabled}) and
$M' = (M - \mbox{$^\bullet G$}) + G^\bullet$. 
\end{definition}

\noindent
Write $M_0 \twoheadrightarrow_N M$ iff there are transitions $t_i\in T$ and markings $M_i\in \powermultiset S$
for $i\mathbin=1,\dots,k$, such that $M_k=M$ and $M_{i{-\!}1} \!\goesto{\{t_{i}\}}_N M_{i}$ for $i\mathbin=1,\dots,k$.
Moreover, $M_0 \!\twoheadrightarrow \goesto{G}$ means that $M_0 \mathbin{\twoheadrightarrow_N} M \mathbin{\goesto{G}_N} M'$
for some $M$ and $M'\!\!$.

\begin{definitionq}{structural conflict}{\cite{GGS11}}
  $N = (S, T, F, M_0, \ell)$ is a \emph{structural conflict net} iff
  $\forall t, u.
    (M_0 \twoheadrightarrow \goesto{\{t, u\}}) \Rightarrow
    \precond{t} \cap \precond{u} = \emptyset$.
\end{definitionq}
Here I restrict myself to structural conflict nets, henceforth simply called \emph{nets}, a class of
Petri nets containing the \emph{safe} Petri nets that are normally used to give semantics to process algebras.

Given a net $N = (S, T, F, M_0, \ell)$, its associated LTS $(\IP,  \Tr, \source,\target,\ell)$
is given by $\IP := \powermultiset{S}$, $\Tr:=\{(M,t)\in\powermultiset{S}\times\Tr\mid \precond{t}\leq M\}$,
$\source(M,t) := M$, $\target(M,t) := (M - \precond{t}) + \postcond{t}$ and $\ell(M,t) := \ell(t)$.
The net $N$ maps to the state $M_0$ in this LTS\@.
A \emph{completeness criterion} on a net is a completeness criterion on its associated LTS\@.
Now $N \models^{CC} \phi$ is defined to hold iff $M_0 \models^{CC} \phi$ in the associated LTS\@.

\subsection{CCS}\label{CCS}

\newcommand{\ca}{a}
\noindent
CCS \cite{Mi90ccs} is parametrised with sets ${\K}$ of \emph{agent identifiers} and $\A$ of \emph{names};
each $X\in\K$ comes with a defining equation \plat{$X \stackrel{{\it def}}{=} P$} with $P$ being a CCS expression as defined below.
\href{https://en.wikipedia.org/wiki/List_of_mathematical_symbols_by_subject#Set_operations}{\raisebox{0pt}[10pt]{$Act := \A\dcup\bar\A\dcup \{\tau\}$}}
is the set of {\em actions}, where $\tau$ is a special \emph{internal action}
and $\bar{\A} := \{ \bar{\ca} \mid \ca \in \A\}$ is the set of \emph{co-names}.
Complementation is extended to $\bar\A$ by setting $\bar{\bar{\mbox{$\ca$}}}=\ca$.
Below, $\ca$ ranges over $\A\cup\bar\A$, $\alpha$ over $Act$, and $X,Y$ over $\K$.
A \emph{relabelling} is a function $f\!:\A\mathbin\rightarrow \A$; it extends to $Act$ by
$f(\bar{\ca})\mathbin=\overline{f(\ca)}$ and $f(\tau):=\tau$.
The set $\cT_{\rm CCS}$ of CCS expressions or \emph{processes} is the smallest set including:
\begin{center}
\begin{tabular}{@{}lll@{}}
$\sum_{i\in I}\alpha_i.P_i$ & for $I$ an index set, $\alpha_i\mathbin\in Act$ and $P_i\mathbin\in\cT_{\rm CCS}$ & \emph{guarded choice} \\
$P|Q$ & for $P,Q\mathbin\in\cT_{\rm CCS}$ & \emph{parallel composition}\\
$P\backslash L$ & for $L\subseteq\A$ and $P\mathbin\in\cT_{\rm CCS}$ & \emph{restriction} \\
$P[f]$ & for $f$ a relabelling and $P\mathbin\in\cT_{\rm CCS}$ & \emph{relabelling} \\
$X$ & for $X\in\K$ & \emph{agent identifier}\\
\end{tabular}
\end{center}
The process $\sum_{i\in \{1,2\}}\alpha_i.P_i$ is often written as $\alpha_1.P_1 + \alpha_2.P_2$,
and $\sum_{i\in \emptyset}\alpha_i.P_i$ as ${\bf 0}$.
The semantics of CCS is given by the transition relation
$\mathord\rightarrow \subseteq \cT_{\rm CCS}\times Act \color{red} \times \Pow(\Ce) \color{black} \times\cT_{\rm CCS}$,
where transitions \plat{$P\goto[C]{\alpha}Q$} are derived from the rules of \autoref{tab:CCS}.
\begin{table*}[t]
\caption{Structural operational semantics of CCS}
\label{tab:CCS}
\normalsize
\begin{center}
  \newcommand\mylabel\weg
  \renewcommand\eta\alpha
  \renewcommand\ell\alpha
\framebox{$\begin{array}{c@{\quad}c@{\qquad}c}
&\sum_{i\in I}\alpha_i.P_i \goto[\{\varepsilon\}]{\alpha_j} P_j\makebox[20pt][l]{~~~~($j\in I$)} \\[3ex]
\displaystyle\frac{P\goto[C]{\eta} P'}{P|Q \goto[\Left\cdot C]{\eta} P'|Q} \mylabel{Par-l}&
\displaystyle\frac{P\goto[C]{\ca} P' ,~ Q \goto[D]{\bar{\ca}} Q'}{P|Q \goto[\Left\cdot C \;\cup\; \R\cdot D]{\tau} P'| Q'} \mylabel{Comm}&
\displaystyle\frac{Q\goto[D]{\eta} Q'}{P|Q \goto[\R\cdot D]{\eta} P|Q'} \mylabel{Par-r}\\[4ex]
\displaystyle\frac{P \goto[C]{\ell} P'}{P\backslash L \goto[C]{\ell}P'\backslash L}~~(\ell,\bar{\ell}\not\in L) \mylabel{Res}&
\displaystyle\frac{P \goto[C]{\ell} P'}{P[f] \goto[C]{f(\ell)} P'[f]} \mylabel{Rel} &
\displaystyle\frac{P \goto[C]{\alpha} P'}{X\goto[C]{\alpha}P'}~~(X \stackrel{{\it def}}{=} P) \mylabel{Rec}
\end{array}$}
\vspace{-2ex}
\end{center}
\end{table*}
Ignoring the labels {\color{red}$C\in \Pow(\Ce)$} for now, such a transition indicates that process
$P$ can perform the action $\alpha\in Act$ and transform into process $Q$.
The process $\sum_{i\in I}\alpha_i.P_i$ performs one of the actions $\alpha_j$ for $j\in I$ and subsequently acts as $P_j$.
The parallel composition $P|Q$ executes an action from $P$, an action from $Q$, or a synchronisation
between complementary actions $c$ and $\bar{c}$ performed by $P$ and $Q$, resulting in an internal action $\tau$. 
The restriction operator $P \backslash L$ inhibits execution of the actions from $L$ and their complements. 
The relabelling $P[f]$ acts like process $P$ with all labels $\alpha$ replaced by $f(\alpha)$.
Finally, the rule for agent identifiers says that an agent $X$ has the same transitions as the body $P$ of its defining equation.
The standard version of CCS \cite{Mi90ccs} features a \emph{choice} operator $\sum_{i\in I}P_i$; here
I use the fragment of CCS that merely features guarded choice.

The second label of a transition indicates the set of (parallel) \emph{components} involved in
executing this transition. The set $\Ce$ of components is defined as $\{\Left,\R\}^*$, that is, the set
of strings over the indicators $\Left$eft and $\R$ight, with $\varepsilon\mathbin\in\Ce$ denoting the empty string
and $\textsc{d}\cdot C := \{\textsc{d}\sigma \mid \sigma\mathbin\in C\}$ for $\textsc{d}\mathbin\in\{\Left,\R\}$ and
$C\mathbin\subseteq \Ce\!$.

\begin{example}{CCS transitions}
The CCS process $P:=(X|\bar a.{\bf 0})|\bar a.b.{\bf 0}$ with $X\stackrel{{\it def}}{=} a.X$ has as outgoing transitions
$P \goto[\{\Left\Left\}]a P$,
$P \goto[\{\Left\Left,\Left\R\}]\tau (X|{\bf 0})|\bar a.b.{\bf 0}$,
~$P \goto[\{\Left\R\}]{\bar a} (X|{\bf 0})|\bar a.b.{\bf 0}$,
~$P \goto[\{\Left\Left,\R\}]\tau (X|\bar a.{\bf 0})|b.{\bf 0}$~ and
~$P \goto[\{\R\}]{\bar a} (X|\bar a.{\bf 0})|b.{\bf 0}$.
\end{example}
These components stem from Victor Dyseryn [personal communication] and were
introduced in \cite{vG19c}. They were not part of the standard semantics of CCS \cite{Mi90ccs},
which can be retrieved by ignoring them.

The LTS of CCS is $(\cT, \Tr, \source,\target,\ell)$, with
$\Tr= \{(P,\alpha,C,Q) \mid P\goto[C]{\alpha}Q\}$, \mbox{$\ell(P,\alpha,C,Q)=\alpha$},
$\source(P,\alpha,C,Q)=P$ and $\target(P,\alpha,C,Q)=Q$.
Employing this interpretation of CCS, one can pronounce judgements $P\models^{CC}\phi$ for CCS processes $P$.

\subsection{Labelled Transition Systems with Concurrency}

\newcommand{\conc}{\smile}
\newcommand{\aconc}{\smile}
\newcommand{\naconc}{\nconc}
\begin{definition}{LTSC}
  A \emph{labelled transition system with concurrency} (LTSC) is a tuple $(\IP, \Tr, \source,\target,\ell,\aconc)$
  consisting of a LTS $(\IP, \Tr, \source,\target,\ell)$ and a \emph{concurrency relation}
  ${\aconc} \subseteq \Tr \times \Tr$, such that:\vspace{-1ex}
  \begin{equation}\label{irreflexivity}
  \mbox{$t \naconc t$ for all $t \in\Tr$,}\vspace{-3ex}
  \end{equation}
  \begin{equation}\label{closure}\begin{minipage}{5.1in}{
  if $t\in\Tr$ and $\pi$ is a path from $\source(t)$ to $s\in \IP$ such that $t \aconc v$ for
  all transitions $v$ occurring in $\pi$, then there is a $u\in\Tr$ such that $\source(u)=s$,
  $\ell(u)=\ell(t)$ and $t \naconc u$.}
  \end{minipage}\end{equation}
\end{definition}
Informally, $t\aconc v$ means that the transition $v$ does not interfere with $t$, in the sense that
it does not affect any resources that are needed by $t$, so that in a state where $t$ and $v$ are
both possible, after doing $v$ one can still do a future variant $u$ of $t$.

LTSCs were introduced in \cite{vG19}, although there the model is more general
on various counts. I do not need this generality in the present paper. In particular, I only need
symmetric concurrency relations $\conc$; in \cite{vG19} $\conc$ is not always symmetric, and denoted
\mbox{$\smile\hspace{-.95ex}\raisebox{2.5pt}{$\scriptscriptstyle\bullet$}$}.

The LTS associated with CCS can be turned into an LTSC by defining $(P,\alpha,C,P') \conc (Q,\beta,D,Q')$
iff $C\cap D = \emptyset$, that is, two transitions are concurrent iff they stem from disjoint sets
of components \cite{GH19,vG19c}.

\begin{example}{CCS transitions concurrency}
Let the 5 transitions from \ex{CCS transitions} be $t$, $u$, $v$, $w$ and $x$, respectively.
Then $t \nconc w$ because these transitions share the component $\Left\Left$. Yet $v \conc w$.
\end{example}

The LTS associated with a Petri net can be turned into an LTSC by defining $(M,t) \conc (M',u)$ iff
$\precond{t} \cap \precond{u} = \emptyset$, i.e., the two LTS-transitions stem from net-transitions
that have no preplaces in common.

Naturally, an LTSC can be turned into a LTS, and further into a Kripke structure, by forgetting $\smile$.

\section{Progress, Justness and Fairness}\label{sec:completeness criteria}

In this section I define completeness criteria $CC \in \{{\it SF}(\Tsk),{\it WF}(\Tsk),J, {\it Pr},\top
\mid \Tsk \in \Pow(\Pow(\Tr))\}$ on LTSs $(\IP, \Tr, \source,\target,\ell)$, to be used in\pagebreak[3]
judgements $P \models^{CC} \phi$, for $P \in \IP$ and $\phi$ an {\LTLX} formula. These
criteria are called \emph{strong fairness} (\emph{SF}), \emph{weak fairness} (\emph{SF}), both
parametrised with a set $\Tsk\subseteq \Pow(\Tr)$ of \emph{tasks}, \emph{justness} ($J$),
\emph{progress} (\emph{Pr}) and the \emph{trivial} completeness criterion ($\top$).
Justness is merely defined on LTSCs.
I confine myself to criteria that hold finite paths ending within a transition to be incomplete.

Reading \ex{beer}, one could find it unfair that Bart might never get a beer.
Strong and weak \emph{fairness} are completeness criteria that postulate that Bart will get a beer,
namely by ruling out as incomplete the infinite paths in which he does not.
They can be formalised by introducing a set $\Tsk$ of \emph{tasks}, each being a set of
transitions (in an LTS or Kripke structure). 
\begin{definitionq}{fairness}{\cite{GH19}}
A task $T \in \Tsk$ is \emph{enabled} in a state $s$ iff $s$ has an outgoing transition from $T$.
It is \emph{perpetually enabled} on a path $\pi$ iff it is enabled in every state of $\pi$.
It is \emph{relentlessly enabled} on $\pi$, if each suffix of $\pi$ contains a state
in which it is enabled.\footnote{This is the case if the task is enabled in infinitely many states
of $\pi$, in a state that occurs infinitely often in $\pi$, or in the last state of a finite $\pi$.}
It \emph{occurs} in $\pi$ if $\pi$ contains a transition $t\in T$.

A path $\pi$ is \emph{weakly fair} if, for every suffix $\pi'$ of $\pi$,
each task that is perpetually enabled on $\pi'$, occurs in $\pi'$.
It is \emph{strongly fair} if, for every suffix $\pi'$ of $\pi$,
each task that is relentlessly  enabled on $\pi'$, occurs~in~$\pi'$.
\end{definitionq}
As completeness criteria, these notions take only the fair paths to be complete.
In \ex{beer} it suffices to have a task ``Bart gets a beer'', consisting of the three transitions
leading to the $B$ state. Now in any path in which Bart never gets a beer this task is perpetually
enabled, yet never taken. Hence weak fairness suffices to rule out such paths.
We have $\mathcal{D} \models^{{\it WF}(\Tsk)} {\bf F}B$.

\emph{Local fairness} \cite{GH19} allows the tasks $\Tsk$ to be declared on an ad hoc basis for the
application at hand. On this basis one can call it unfair if Bart doesn't get a beer,
without requiring that Cameron should get a beer as well. \emph{Global fairness}, on the other hand,
distils the tasks of an LTS in a systematic way out of the structure of a formalism, such as
pseudocode, process algebra or Petri nets, that gave rise to the LTS\@.
A classification of many ways to do this, and thus of many notions of strong and weak fairness,
appears in \cite{GH19}. In \emph{fairness of directions} \cite{Fr86}, for instance, each
transition in an LTS is assumed to stem from a particular \emph{direction}, or \emph{instruction},
in the pseudocode that generated the LTS; now each direction represents a task, consisting of all
transitions derived from that direction.

In \cite{GH19} the assumption that a system will never stop when there are
transitions to proceed is called \emph{progress}. In \ex{Bart alone} it takes a progress
assumption to conclude that Bart will get his beer. Progress fits the default completeness criterion
introduced before, i.e., $\models^{\it Pr}$ is the same as $\models$. Not (even) assuming progress
can be formalised by the trivial completeness criterion $\top$ that declares all paths to be
complete. Naturally, $\mathcal{E} \not\models^\top {\bf F}B$.

Completeness criterion $D$ is called \emph{stronger} than criterion $C$ if it rules out
more paths as incomplete. So $\top$ is the weakest of all criteria, and, for any given collection
$\Tsk$, strong fairness is stronger than weak fairness. When assuming that each transition occurs in
at least one task---which can be ensured by incorporating a default task consisting of all
transitions---progress is weaker than weak fairness.

\emph{Justness} \cite{GH19} is a strong form of progress, defined on LTSCs.
\begin{definition}{justness}
A path $\pi$ is \emph{just} if for each transition $t$ with its source state $s := \source(t)$ occurring on $\pi$,
the suffix of $\pi$ starting at $s$ contains a transition $u$ with $t \nconc u$.
\end{definition}
\begin{example}{CCS justness}
The infinite path $\pi$ that only ever takes transition $t$ in \ex{CCS transitions}/\ref{ex:CCS transitions concurrency} is unjust.
Namely with transition $v$ in the r\^ole of the $t$ from \df{justness}, $\pi$ contains no transition $y$ with $v \nconc y$.
\end{example}
Informally, the only reason for an enabled transition not to occur, is that one of its resources is
eventually used for some other transition. In \ex{Bart separated} for instance, the orders of Alice
and Cameron are clearly concurrent with the one of Bart, in the sense that they do not compete for
shared resources. Taking $t$ to be the transition in which Bart gets his beer, any path in which
$t$ does not occur is unjust. Thus $\mathcal{F} \models^J {\bf F}B$.

For most choices of $\Tsk$ found in the literature, weak fairness is a strictly stronger
completeness criterion than justness. In \ex{beer}, for instance, the path in which Bart does not
get a beer is just. Namely, any transition $u$ giving Alice or Cameron a beer competes for the same
resource as the transition $t$ giving Bart a beer, namely the attention of the barman.
Thus $t \nconc u$, and consequently $\mathcal{D} \not\models^J {\bf F}B$.

\section{Reactive Temporal Logic}\label{sec:LTL}

Standard treatments of temporal logic \cite{Pnueli77,HuthRyan04} are adequate for \emph{closed systems},
having no run-time interactions with their environment. However, they fall short for \emph{reactive systems},
interacting with their environments through synchronisation of actions.
\newcommand{\VM}{{\it VM}}

\begin{example}{pretzel}
Consider a vending machine that accepts a coin $c$ and
produces a pretzel $p$. We assume that accepting the coin requires
cooperation from the user/environment, but producing the pretzel does not.
A CCS specification is\vspace{-1ex} $$\VM=c.p.\VM\;.$$
In standard {\LTLX} (assuming progress) we have $\VM \models {\bf G}(c\Rightarrow {\bf F}p)$.
This formula says that whenever a coin is inserted, eventually a
pretzel is produced. This formula is intuitively true indeed.
But we also have $\VM \models {\bf G}(p\Rightarrow {\bf F}c)$.
This formula says that whenever a pretzel is produced, eventually a new
coin will be inserted. This formula is intuitively false.
This example shows that standard {\LTLX} is not suitable to correctly describe
the behaviour of this vending machine.
\end{example}
For this reason I here introduce \emph{reactive} {\LTLX}.
The syntax and semantics are unchanged, except that I use a validity relation $\models_B$
that is parametrised with a set $B\subseteq A$ of \emph{blockable} actions.
Here $A$ is the set of all observable actions of the LTS on which {\LTLX} is interpreted.
The intuition is that actions $b\in B$ may be blocked by the environment, but actions
$a \in A {\setminus} B$ may not. The relation $\models_B$ can be used to formalise
the assumption that the actions in $A {\setminus} B$ are not under the control of the user of the
modelled system, or that there is an agreement with the user not to block them. Either way, it is a
disclaimer on the wrapping of our temporal judgement, that it is valid only when applying the
involved distributed system in an environment that may block actions from $B$ only. The hidden
action $\tau$ may never be blocked.

The subscript $B$ modifies the default completeness criterion, to call a path complete iff it is
either infinite or ends in a state of which all outgoing transitions have a label from $B$.
Note that the standard {\LTLX} interpretation $\models$ is simply
$\models_\emptyset$, obtained by taking the empty set of blocking actions.

In \ex{pretzel} one takes $B=\{c\}$. This choice of $B$ says that the environment may
block the action $c$, namely by not inserting a coin; however, the environment may not block $p$.  
As intuitively expected, we have
$\VM \models_B {\bf G}(c\Rightarrow {\bf F}p)$ but $\VM \not\models_B {\bf G}(p\Rightarrow {\bf F}c)$.

Naturally, reactive {\LTLX} can also be combined with a non-default completeness criterion, as
discussed in Sections~\ref{sec:Kripke}--\ref{sec:completeness criteria}. When writing $P \models^{CC}_B\phi$
the modifier $B$ adapts the default completeness criterion by declaring certain finite paths
complete, and the modifier $CC\neq \top$ adapts it by declaring some
infinite paths incomplete. In the presence of the modifier $B$, \df{justness} and the first sentence
of \df{fairness} are adapted as follows:
\begin{definition}{Bjustness}
A path $\pi$ is \emph{just} (or \emph{$B$-just}) if for each transition $t\in\Tr$ with
$\ell(t)\notin B$ and its source state $s := \source(t)$ occurring on $\pi$,
the suffix of $\pi$ starting at $s$ contains a transition $u$ with $t \nconc u$.
\end{definition}
Note that it doesn't matter whether $\ell(u)\in B$ or not.
\begin{definition}{Bfairness}
A task $T \in \Tsk$ is \emph{enabled} in a state $s$ iff $s$ has an outgoing transition $t\in T$
with $\ell(t)\notin B$.
\end{definition}
The above completes the formal definition of the validity of temporal judgements $P \models^{CC}_B \phi$
with $\phi$ an {\LTLX} formula, $B \subseteq A$, and either
\begin{itemize}
\vspace{-3pt}
\item $CC={\it Pr}$ and $P$ a state in an LTS, a CCS expression or a Petri net,
\vspace{-1ex}
\item $CC={\it J}$ and $P$ a state in an LTSC, a CCS expression or a Petri net,
\vspace{-1ex}
\item $CC={\it WF}(\Tsk)$ or ${\it SF}(\Tsk)$ and $P$ a state in an LTS
   $(\IP, \Tr, \source,\target,\ell)$ with $\Tsk\in\Pow(\Pow(\Tr))$,
   or $P$ a CCS expression or Petri net with associated LTS
   $(\IP, \Tr, \source,\target,\ell)$ and $\Tsk\in\Pow(\Pow(\Tr))$.%
\vspace{-3pt}
\end{itemize}
Namely, in case $P$ is a state in an LTS, it is also a state in the associated Kripke structure $K$.
Moreover, $B$ and $CC$ combine into a single completeness criterion ${\it BC}$ on that LTS, which
translates as a completeness criterion ${\it BC}$ on $K$. Now \df{validity} tells whether
$P\models^{\it BC} \phi$ holds.

In case $CC=J$ and $P$ a state in an LTSC,
$B$ and $J$ combine into a single completeness criterion ${\it BJ}$ on that LTSC, which is also a
completeness criterion on the associated LTS; now proceed as above.

In case $P$ is a Petri net or CCS expression, first translate it into a state in an LTS or LTSC,
using the translations at the end of Sections~\ref{sec:nets} or \ref{CCS}, respectively, and proceed as above.

Temporal judgements $P \models^{CC}_B \phi$, as introduced above, are not limited to the case that
$\phi$ is an LTL formula. In \Sec{CTL} I will show that allowing $\phi$ to be a CTL formula instead
poses no additional complications, and I expect the same to hold for other temporal logics.

Judgements $P \models^{CC}_B \phi$ get stronger (= less likely true) when the completeness criterion
$CC$ is weaker, and the set $B$ of blockable actions larger.

Most concepts of reactive temporal logic introduced above stem from
\cite{GH15a}.  The main novelty contributed here is the annotated satisfaction relation
$\models^{CC}_B$. In \cite{GH15a} we simply wrote $\models$, expecting $CC$ and $B$ to be determined
once and for all in a given paper or application.  Requirement specifications in which different
values for $B$ are combined, such as FS\hspace{1pt}1--2 in \Sec{formalising FS}, were not foreseen there.

\section{The Mutual Exclusion Problem and its History}\label{sec:history}

The mutual exclusion problem was presented by Dijkstra in \cite{Dijk65} and formulated as follows:
\begin{quote}
   ``To begin, consider $N$ computers, each engaged in a
process which, for our aims, can be regarded as cyclic. In
each of the cycles a so-called ``critical section'' occurs and
the computers have to be programmed in such a way that
at any moment only one of these $N$ cyclic processes is in
its critical section. In order to effectuate this mutual
exclusion of critical-section execution the computers can
communicate with each other via a common store. Writing
a word into or nondestructively reading a word from this
store are undividable operations; i.e., when two or more
computers try to communicate (either for reading or for
writing) simultaneously with the same common location,
these communications will take place one after the other,
but in an unknown order.''
\end{quote}
Dijkstra proceeds to formulate a number of requirements
that a solution to this problem must satisfy, and then
presents a solution that satisfies those requirements. 
The most central of these are:
\begin{itemize}
\vspace{-3pt}
\item (\emph{Safety}) ``no two computers can be in their critical section simultaneously'', and
\vspace{-3pt}
\item (\emph{Dijkstra's Liveness}) If at least one computer intends to enter its critical section,
then at least one ``will be allowed to enter its critical section in due time''.
\vspace{-3pt}
\end{itemize}
Two other important requirements formulated by Dijkstra are
\begin{itemize}
\vspace{-3pt}
\item (\emph{Speed independence}) ``(b) Nothing may be assumed about the relative speeds of the $N$
  computers'',%
\vspace{-3pt}
\item and (\emph{Optionality}) ``(c) If any of the computers is stopped well outside its
critical section, this is not allowed to lead to potential
blocking of the others.''
\end{itemize}
A crucial assumption is that each computer, in each cycle, spends only a finite amount of time in its
critical section. This is necessary for the correctness of any mutual exclusion protocol.

For the purpose of the last requirement one can partition each cycle into a \emph{critical section}, a
\emph{noncritical section} (in which the process starts), an \emph{entry protocol} between the
noncritical and the critical section, during which a process prepares for entry in negotiation with the
competing processes, and an \emph{exit protocol}, that comes right after the critical section and
before return to the noncritical section. Now ``well outside its critical section'' means in the
noncritical section. 
Requirement (c) can equivalently be stated as admitting the possibility that a process chooses to
remain forever in its noncritical section, without applying for entry in the critical section ever again.

Knuth \cite{Knuth66} proposes a strengthening of Dijkstra's liveness requirement, namely
\begin{itemize}
\item (\emph{Liveness}) If a computer intends to enter its critical section,
then it will be allowed to enter in due time.
\end{itemize}
He also presents a solution that is shown to satisfy this requirement, as well as Dijkstra's requirements.%
\footnote{It can be argued, however, that Knuth's mutual exclusion protocol is correct only when
  making certain assumptions on
  the hardware on which it will be running \cite{vG18c}; the same applies to all other mutual
  exclusion protocols mentioned in this section. This matter is not addressed in the present paper.
  However, the material presented in \Sec{formalising mutex} paves the way for discussing it.}
Henceforth I define a correct solution of the mutual exclusion problem as one that satisfies both
safety and liveness, as formulated above, as well as optionality. I sometimes speak of ``speed
independent mutual exclusion'' when also insisting on requirement (b) above.

The special case of the mutual exclusion problem for two processes ($N=2$) was presented by Dijkstra in
\cite{EWD35}, three years prior to \cite{Dijk65}. There Dijkstra presented a solution found by
T.J. Dekker in 1959, and shows that it satisfies all requirements of \cite{Dijk65}.
Although not explicitly stated in \cite{EWD35}, the arguments given therein imply straightforwardly
that Dekker's solution also satisfy the liveness requirement above.

Peterson \cite{Pet81} presented a considerable simplification of Dekker's algorithm that satisfies
the same correctness requirements.
Many other mutual exclusion protocols appear in the literature, the most prominent being
Lamport's bakery algorithm \cite{bakery} and Szyma\'nski's mutual exclusion algorithm \cite{Szy88}.
These guarantee some additional correctness criteria besides the ones discussed above.

\section{Fair Schedulers}\label{sec:FS}

\makebox[0pt][l]{\raisebox{-13ex}[0pt][0pt]{\small FS\hspace{1pt}1}}%
\makebox[0pt][l]{\raisebox{-30ex}[0pt][0pt]{\small FS\hspace{1pt}2}}%
\makebox[0pt][l]{\raisebox{-36ex}[0pt][0pt]{\small FS\hspace{1pt}3}}%
In \cite{GH15b} a \emph{fair scheduler} is defined as
\begin{quote}
``a reactive system with two input channels: one on which it can receive requests $r_1$ from its
  environment and one on which it can receive requests $r_2$. We allow the scheduler to be too busy
  shortly after receiving a request $r_i$ to accept another request $r_i$ on the same channel.
  However, the system will always return to a state where it remains ready to accept the next
  request $r_i$ until $r_i$ arrives. In case no request arrives it remains ready forever. The
  environment is under no obligation to issue requests, or to ever stop issuing requests.  Hence for
  any numbers $n_1$ and $n_2\in\IN\cup\{\infty\}$ there is at least one run of the system in which
  exactly that many requests of type $r_1$ and $r_2$ are received.

  Every request $r_i$ asks for a task $t_i$ to be executed.  The crucial property of the fair
  scheduler is that it will eventually grant any such request. Thus, we require that in any run of
  the system each occurrence of $r_i$ will be followed by an occurrence of $t_i$.''

``We require that in any partial run of the scheduler there may not be more occurrences of $t_i$
  than of $r_i$, for $i=1,2$.

  The last requirement is that between each two occurrences of $t_i$ and $t_j$ for $i,j\in\{1,2\}$
  an intermittent activity $e$ is scheduled.''
\end{quote}
\makebox[0pt][l]{\raisebox{6ex}[0pt][0pt]{\small FS\hspace{1pt}4}}%
This fair scheduler serves two clients, but the concept generalises smoothly to $N$ clients.

The intended applications of fair schedulers are for instance in operating systems, where multiple
application processes compete for processing on a single core, or radio broadcasting stations, where 
the station manager needs to schedule multiple parties competing for airtime. In such cases
each applicant must get a turn eventually. The event $e$ signals the end of the time slot allocated
to an application process on the single core, or to a broadcast on the radio station.

Fair schedulers occur (in suitable variations) in many distributed systems.  Examples are
\emph{First in First out}\footnote{Also known as First Come First Served (FCFS)},
\emph{Round Robin}, and
\emph{Fair Queueing} 
scheduling algorithms\footnote{\url{http://en.wikipedia.org/wiki/Scheduling_(computing)}}
as used in network routers~\cite{rfc970,Nagle87} and operating systems~\cite{Kleinrock64}, or the
\emph{Completely Fair Scheduler},\footnote{\url{http://en.wikipedia.org/wiki/Completely_Fair_Scheduler}}
which is the default scheduler of the Linux kernel since version 2.6.23.

Each action $r_i$, $t_i$ and $e$ can be seen as a communication
between the fair scheduler and one of its clients. In a reactive system such communications will take place
only if both the fair scheduler and its client are ready for it. Requirement FS\hspace{1pt}1 of a fair
scheduler quoted above effectively shifts the responsibility for executing $r_i$ to the client.
The actions $t_i$ and $e$, on the other hand, are seen as the responsibility of the fair scheduler.
We do not consider the possibility that the fair scheduler fails to execute $t_i$ merely because the
client does not collaborate. Hence \cite{GH15b} assumes that the client cannot prevent the actions
$t_i$ and $e$ from occurring. It is furthermore assumed that executing the actions $r_i$, $t_i$
and $e$ takes a finite amount of time only.

A fair scheduler closely resembles a mutual exclusion protocol.
However, its goal is not to achieve mutual exclusion. In most applications, mutual exclusion can be
taken for granted, as it is physically impossible to allocate the single core to multiple applications
at the same time, or the (single frequency) radio sender to multiple simultaneous broadcasts.
Instead, its goal is to ensure that no applicant is passed over forever.

It is not hard to obtain a fair scheduler from a mutual exclusion protocol.
For suppose we have a mutual exclusion protocol $M$, serving two processes $P_i$ ($i=1,2$).
I instantiate the non-critical section of process $P_i$ as patiently awaiting the request $r_i$.
As soon as this request arrives, $P_i$ leaves the noncritical section and starts the entry protocol
to get access to the critical section. The liveness property for mutual exclusion guarantees that
$P_i$ will reach its critical section. Now the critical section consists of scheduling task $t_i$,
followed by the intermittent activity $e$. Trivially, the composition of the two process $P_i$, in
combination with protocol $M$, constitutes a fair scheduler, in that it meets the above four requirements.

One can not quite construct a mutual exclusion protocol from a fair scheduler, due to fact that in a
mutual exclusion protocol leaving the critical section is controlled by the client process.
For this purpose one would need to adapt the assumption that the client of a fair scheduler cannot block
the intermittent activity $e$ into the assumption that the client can postpone this action, but for
a finite amount of time only. In this setting one can build a mutual exclusion protocol, serving two
processes $P_i$ ($i=1,2$), from a fair scheduler $F$. Process $i$ simply issues request $r_i$ at
$F$ as soon as it has left the non-critical section, and when $F$ communicates the action $t_i$,
Process $i$ enters its critical section. Upon leaving its critical section, which is assumed to
happen after a finite amount of time, it participates in the synchronisation $e$ with
$F$. Trivially, this yields a correct mutual exclusion protocol.

\newpage

\section[Formalising the Requirements for Fair Schedulers in Reactive LTL]
        {Formalising the Requirements for Fair Schedulers in Reactive {\LTLX}}\label{sec:formalising FS}

The main reason fair schedulers were defined in \cite{GH15b} was to serve as an example
of a realistic class of systems of which no representative can be correctly specified in CCS, or
similar process algebras, or in Petri nets. Proving this impossibility result necessitated a precise
formalisation of the four requirements quoted in \Sec{FS}. Through the provided translations of CCS
and Petri nets into LTSs, a fair scheduler rendered in CCS or Petri nets can be seen as a state $F$
in an LTS over the set $\{r_i,t_i,e \mid i=1,2\}$ of visible actions; all other actions can be
considered internal and renamed into $\tau$.

Let a \emph{partial trace} of a state $s$ in an LTS be the sequence of visible actions encountered on
a path starting in $s$ \cite{vG93}. Now the last two requirements (FS\hspace{1pt}3) and (FS\hspace{1pt}4) of a fair
scheduler are simple properties that should be satisfied by all partial traces $\sigma$ of state $F$:
{\leftmargini 50pt \labelsep 15pt
\begin{itemize}
\item[(FS\hspace{1pt}3)] $\sigma$ contains no more occurrences of $t_i$ than of $r_i$, for $i=1,2$,
\item[(FS\hspace{1pt}4)] $\sigma$ contains an occurrence of $e$ between each two occurrences of $t_i$ and $t_j$ for $i,j\in\{1,2\}$.
\end{itemize}}
\noindent
FS\hspace{1pt}4 can be conveniently rendered in {\LTLX}:
{\leftmargini 50pt \labelsep 15pt
\begin{itemize}
\item[(FS\hspace{1pt}4)]
$F \models {\bf G}\left(t_i \Rightarrow \textcolor{red}{\big(}t_i {\bf U} 
 \big((\neg t_1 \wedge \neg t_2) {\bf W} e\big)\textcolor{red}{\big)}\right)$
\end{itemize}}
\noindent
for $i\in\{1,2\}$.
Here the \emph{weak until} modality $\psi {\bf W} \phi$ is syntactic sugar for ${\bf G}\psi \vee (\psi {\bf U} \phi)$.
If I hadn't lost the {\bf X} modality, I could write {\bf X} for $t_i {\bf U}$ in the above formula;
on Kripke structures distilled from LTSs the meaning is the same. The formula in FS\hspace{1pt}4 is of a kind
where the meaning of $\models_B^{CC}$ is independent of $B$ and $CC$. This follows from the fact
that FS\hspace{1pt}4 merely formulates a property that should hold for all partial runs.
Hence one need not worry about which $B$ and $CC$ to employ here.

Unfortunately, FS\hspace{1pt}3 cannot be formulated in {\LTLX}, due to the need to keep count of the
difference in the number of $r_i$ and $t_i$ actions encountered on a path. However, one could strengthen FS\hspace{1pt}3 into
{\leftmargini 50pt \labelsep 15pt
\begin{itemize}
\item[(FS\hspace{1pt}3$'$)] $\sigma$ contains an occurrence of $r_i$ between each two occurrences of $t_i$, and
  prior to the first occurrence of $t_i$, for $i\in\{1,2\}$.
\end{itemize}}
\noindent
This would restrict the class of acceptable fair schedulers, but keep the most interesting examples.
Consequently, the impossibility result from \cite{GH15b} applies to this modified class as well.
FS\hspace{1pt}3 can be rendered in {\LTLX} in the same style as FS\hspace{1pt}4:
{\leftmargini 50pt \labelsep 15pt
\begin{itemize}
\item[(FS\hspace{1pt}3$'$)] $F \models  \textcolor{red}{\big(}(\neg t_i) {\bf W} r_i \textcolor{red}{\big)}
  \wedge {\bf G}\left(t_i \Rightarrow
   \textcolor{red}{\big(}t_i {\bf U} \big((\neg t_i) {\bf W} r_i\big) \textcolor{red}{\big)}\right)$
\end{itemize}}
\noindent
for $i\in\{1,2\}$.

Requirement FS\hspace{1pt}2 involves a quantification over all complete runs of the system, and thus depends
on the completeness criterion $CC$ employed. It can be formalised as
{\leftmargini 50pt \labelsep 15pt
\begin{itemize}
\item[(FS\hspace{1pt}2)] $F \models_B^{CC}{\bf G}(r_i \Rightarrow {\bf F}t_i)$
\end{itemize}}
\noindent
for $i\in\{1,2\}$, where $B=\{r_1,r_2\}$.
The set $B$ should contain $r_1$ and $r_2$, as these actions are supposed to be under the control
of the users of a fair scheduler. However, actions $t_1$, $t_2$ and $e$ should not be in $B$, as they
are under the control of the scheduler itself.
In \cite{GH15b}, the completeness criterion employed is justness, so the above formula with $CC:=J$
captures the requirement on the fair schedulers that are shown in \cite{GH15b} not to exist in CCS
or Petri nets. However, keeping $CC$ a variable allows one to pose to the question under which
completeness criterion a fair scheduler \emph{can} be rendered in CCS\@. Naturally, it needs to be
a stronger criterion than justness. In \cite{GH15b} it is shown that weak fairness suffices.

FS\hspace{1pt}2 is a good example of a requirement that can \emph{not} be rendered correctly in standard LTL\@.
Writing $F \models^{CC}{\bf G}(r_i \Rightarrow {\bf F}t_i)$ would rule out the complete runs of $F$
that end because the user of $F$ never supplies the input $r_j \in B$.  The CCS process\vspace{-2ex}
$$F \stackrel{{\it def}}{=} r_1.r_2.t_1.e.t_2.e.F$$
for instance satisfies this formula, as well as FS\hspace{1pt}3 and 4; yet it does not satisfy requirement FS\hspace{1pt}2.
Namely, the path consisting of the $r_1$-transition only is complete, since it ends in a state
of which the only outgoing transition has the label $r_2\in B$. Yet on this path $r_1$ is not followed by $t_1$.

Requirement FS\hspace{1pt}1 is by far the hardest to formalise. In \cite{GH15b} two formalisations are shown to
be equivalent: one involving a coinductive definition of $B$-just paths that exploits the syntax of
CCS, and the other requiring that requirements FS\hspace{1pt}2--4 are preserved under putting an input
interface around process~$F$. The latter demands that also
$\widehat F:= (I_1\,|\,F[f]\,|\,I_2)\backslash \{c_1,c_2\}$ should satisfy FS\hspace{1pt}2--4;\vspace{2pt}
here $f$ is a relabelling with $f(r_i)=c_i$, $f(t_{i})=t_{i}$ and $f(e)=e$ for $i=1,2$,
and \plat{$I_{i}\stackrel{\it def}{=} r_i.\bar{c_i}.I_{i}$} for $i\in\{1,2\}$.

A formalisation of FS\hspace{1pt}1 on Petri nets also appears in \cite{GH15b}: each complete path $\pi$ with
only finitely many occurrences of $r_i$ should contain a state (= marking) $M$, such that there is a
transition $v$ with $\ell(v)=r_i$ and $\precond{v}\leq M$, and for each transition $u$ that occurs
in $\pi$ past $M$ one has $\precond{v} \cap \precond{u} = \emptyset$.

When discussing proposals for fair schedulers by others, FS\hspace{1pt}1 is
the requirement that is most often violated, and explaining why is not always easy.

In reactive {\LTLX}, this requirement is formalised as
{\leftmargini 50pt \labelsep 15pt
\begin{itemize}
\item[(FS\hspace{1pt}1)] $F \models_{B\setminus\{r_i\}}^J {\bf GF} r_i$
\end{itemize}}
\phantomsection\label{FS1}
\noindent
for $i\in\{1,2\}$, or $F \models_{B\setminus\{r_i\}}^{\it CC} {\bf GF} r_i$ if one wants to discuss the
completeness criterion $CC$ as a parameter. The surprising element in this temporal judgement is the subscript
${B\setminus\{r_i\}} = \{r_{3-i}\}$, which contrasts with the assumption that requests are under
the control of the environment.
FS\hspace{1pt}1 says that, although we know that there is no guarantee that user $i$ of $F$ will ever issue
request $r_i$, under the assumption that the user \emph{does} want to make such a request, making
the request should certainly succeed. This means that the protocol itself does not sit in the way of making this request.

The combination of requirements FS\hspace{1pt}1 and 2, which use different sets of blockable actions as a
parameter, is enabled by reactive {\LTLX} as presented here.

The following examples, taken from \cite{GH15b}, show that all the above requirements are necessary for the result from
\cite{GH15b} that fair schedulers cannot be rendered in CCS\@.
\begin{itemize}
\item The CCS process $F_1|F_2$ with \plat{$F_i \stackrel{{\it def}}{=} r_i.t_i.e.F_i$} satisfies 
  FS\hspace{1pt}1, FS\hspace{1pt}2 and FS\hspace{1pt}3$'$. In FS\hspace{1pt}1 and 2 one needs to take ${\it CC} := J$, as progress is not a strong
  enough assumption here.
\item The process $E_1|G|E_2$ with \plat{$E_i \mathbin{\stackrel{{\it def}}{=}} r_i.E_i$} and
  \plat{$G \stackrel{{\it def}}{=} t_1.e.t_2.e.G$} satisfies FS\hspace{1pt}1, 2 and 4,
  again with ${\it CC} \mathbin{:=} J$.
\item The process $E_1|E_2$ satisfies FS\hspace{1pt}1, 3$'$ and 4, again with ${\it CC} \mathbin{:=} J$ in FS\hspace{1pt}1.
\item The process $F_0$ with \plat{$F_0 \stackrel{{\it def}}{=} r_1.t_1.e.F_0 + r_2.t_2.e.F_0$} satisfies FS\hspace{1pt}2--4.
  Here FS\hspace{1pt}2 merely needs ${\it CC} \mathbin{:=} {\it Pr}$, that is, the assumption of progress.
  Furthermore, it satisfies FS\hspace{1pt}1 with ${\it CC} := {\it SF}(\Tsk)$, as long as
  $\textsc{r}_1,\textsc{r}_2 \in \Tsk$. Here $\textsc{r}_i$ is the set of transitions with label $r_i$.
\end{itemize}

{\makeatletter
\let\par\@@par
\par\parshape0
\everypar{}\begin{wrapfigure}[14]{r}{0.24\textwidth}
 \vspace{-1ex}
 \refstepcounter{figure}
 \input{gatekeeper}
 \centerline{\raisebox{1ex}{\box\graph}}
\end{wrapfigure}
\noindent
  The process $X$ given by $X \stackrel{{\it def}}{=} r_1.Y + r_2.Z$,
  ~$Y \stackrel{{\it def}}{=} r_2.t_1.e.Z + t_1.(r_2.e.Z + e.X)$ and
  \plat{$Z \stackrel{{\it def}}{=} r_1.t_2.e.Y + t_2.(r_1.e.Y + e.X)$}, the \emph{gatekeeper},
  is depicted on the right.
The grey shadows represent copies of the states at the opposite end of the diagram, so the
transitions on the far right and bottom loop around.
This process satisfies FS\hspace{1pt}3$'$ and 4, FS\hspace{1pt}2 with ${\it CC} \mathbin{:=} {\it Pr}$,
  and FS\hspace{1pt}1 with ${\it CC} := {\it WF}(\Tsk)$, thereby improving process $F_0$, and
  constituting the best CCS approximation of a fair scheduler seen so far.
  Yet, intuitively FS\hspace{1pt}1 is not ensured at all, meaning that weak
  fairness is too strong an assumption. Nothing really prevents all the choices between $r_2$ and
  any other action $a$ to be made in favour of $a$.
\par}

\section[Formalising Requirements for Mutual Exclusion in Reactive LTL]
        {Formalising Requirements for Mutual Exclusion in Reactive {\LTLX}}\label{sec:formalising mutex}

Define a process $i$ participating in a mutual exclusion protocol to cycle through the stages
\emph{noncritical section}, \emph{entry protocol}, \emph{critical section}, and \emph{exit protocol},
in that order, as explained in \Sec{history}. Modelled as an LTS, its visible actions will be
$\en$, $\lni$, $\ec$ and $\lc$, of entering and leaving its (non)critical section.
Put $\lni$ in $B$ to make leaving the critical section a blockable action.
The environment blocking it is my way of allowing the client process to stay in its noncritical
section forever. This is the manner in which the requirement \emph{Optionality} is captured in reactive temporal logic.
On the other hand, $\ec$ should not be in $B$, for one does not consider the liveness property of a
mutual exclusion protocol to be violated simply because the client process refuses to enter the
critical section when allowed by the protocol. Likewise, $\en$ is not in $B$. Although exiting the
critical section is in fact under control of the client process, it is assumed that it
will not stay in the critical section forever. In the models of this paper this can be simply
achieved by leaving $\lc$ outside $B$. Hence $B := \{\lni \mid i=1,\dots,N\}$.
\newcommand{\act}{\,{\it act}_i}

My first requirement on mutual exclusion protocols $P$ simply says that the actions $\en$, $\lni$,
$\ec$ and $\lc$ have to occur in the right order:
{\leftmargini 50pt \labelsep 15pt
\begin{itemize}
\item[(ME\,1)] $P \models  \begin{array}[t]{r}
  \textcolor{red}{\big(}(\neg \act) {\bf W} \lni \textcolor{red}{\big)}
  \wedge {\bf G}\left(\lni \Rightarrow
   \textcolor{red}{\big(}\lni {\bf U} \big((\neg \act) {\bf W} \ec\big) \textcolor{red}{\big)}\right)
  \wedge {\bf G}\left(\ec \Rightarrow
   \textcolor{red}{\big(}\ec {\bf U} \big((\neg \act) {\bf W} \lc\big) \textcolor{red}{\big)}\right)
  \\ \mbox{}
  \wedge {\bf G}\left(\lc \Rightarrow
   \textcolor{red}{\big(}\lc {\bf U} \big((\neg \act) {\bf W} \en\big) \textcolor{red}{\big)}\right)
  \wedge {\bf G}\left(\en \Rightarrow
   \textcolor{red}{\big(}\en {\bf U} \big((\neg \act) {\bf W} \lni\big) \textcolor{red}{\big)}\right)
\end{array}$
\end{itemize}}
\noindent
for $i=1,\dots,N$. Here $\act := (\lni \vee \ec \vee \lc \vee \en)$.

The second is a formalisation of \textit{Safety}, saying that only one process can be in its critical
section at the same time:
{\leftmargini 50pt \labelsep 15pt
\begin{itemize}
\item[(ME\,2)]
$P \models {\bf G}\left(\ec \Rightarrow \textcolor{red}((\neg \textcolor{red}{\textit{ec}_j}) {\bf W} \lc\textcolor{red})\right)$
\end{itemize}}
\noindent
for $i,j=1,\dots,N$ with $i\neq j$.
Both ME\,1 and ME\,2 would be unaffected by changing $\models$ into $\models^{CC}$ or $\models_B^{CC}$.

Requirement \emph{Liveness} of \Sec{history} can be formalised as
{\leftmargini 50pt \labelsep 15pt
\begin{itemize}
\item[(ME\,3)] $P \models_B^{CC} {\bf G}(\lni \Rightarrow {\bf F}\ec)$
\end{itemize}}
\noindent
Here the choice of a completeness criterion is important.
Finally, the following requirements are similar to \emph{Liveness}, and state that from each section in
the cycle of a Process $i$, the next section will in fact be reached.
In regards to reaching the end of the noncritical section, this should be guaranteed only when
assuming that the process wants to leave it critical section; hence $\lni$ is excepted from $B$.
{\leftmargini 50pt \labelsep 15pt
\begin{itemize}
\item[(ME\,4)] $P \models_B^{CC} {\bf G}(\ec \Rightarrow {\bf F}\lc)$
\item[(ME\,5)] $P \models_B^{CC} {\bf G}(\lc \Rightarrow {\bf F}\en)$
\item[(ME\,6)] $P \models_{B\setminus\mbox{\scriptsize\{\lni\}}}^{CC} {\bf F}\lni
   \wedge {\bf G}(\en \Rightarrow {\bf F}\lni)$
\end{itemize}}
\noindent
for $i=1,\dots,N$.

The requirement \emph{Speed independence} is automatically satisfied for models of mutual
exclusion protocols rendered in any of the formalisms discussed in this paper, as these formalisms
lack the expressiveness to make anything dependent on speed.

The following examples show that none of the above requirements are redundant.
\begin{itemize}
\item The CCS process $F_1|F_2| \cdots|F_N$ with \plat{$F_i \stackrel{{\it def}}{=}
  \lni.\ec.\lc.\en.F_i$} satisfies all requirements, with ${\it CC} := J$, except for ME\,2.
\item The process $R_1|R_2| \cdots|R_N$ with \plat{$R_i \stackrel{{\it def}}{=} \lni.{\bf 0}$}
  satisfies all requirements except for ME\,3.
\item In case $N\mathbin=2$, the process 
   {\color{blue}\it ln$_1$}.{\color{red}\it ec$_1$}.{\color{blue}\it ln$_2$}.{\color{red}\it ec$_2$}.{\bf 0}
   \!+\! {\color{blue}\it ln$_2$}.{\color{red}\it ec$_2$}.{\color{blue}\it ln$_1$}.{\color{red}\it ec$_1$}.{\bf 0}
   satisfies all requirements except for ME\,4.
   The case $N>2$ is only notationally more cumbersome.
   In the same spirit one finds counterexamples failing only on ME\,5, or on the second conjunct of ME\,6.
\item The process $\bf 0$ satisfies all requirements except for the first conjunct of ME\,6.
\item In case $N\mathbin=1$, the process $X$ with
   \plat{$X \stackrel{{\it def}}{=} \mbox{\color{red}\it lc$_1$}.
   \mbox{\color{red}\it ec$_1$}.\mbox{\color{red}\it lc$_1$}.\mbox{\color{blue}\it en$_1$}.
   \mbox{\color{blue}\it ln$_1$}.X$} satisfies all requirements but ME\,1.
\end{itemize}
{\makeatletter
\let\par\@@par
\par\parshape0
\everypar{}\begin{wrapfigure}[11]{r}{0.313\textwidth}
 \vspace{-1ex}
 \input{gatekeeper2}
  \centerline{\raisebox{1ex}{\box\graph}}
 \end{wrapfigure}
\noindent
  The process $X$, a gatekeeper variant, given by 
  \plat{$X \stackrel{{\it def}}{=} \lni[1].Y + \lni[2].Z$},\\[2pt]
  \plat{$Y \stackrel{{\it def}}{=} \lni[2].\ec[1].\lc[1].\en[1].Z + \ec[1].(\lni[2].\lc[1].\en[1].Z +
  \lc[1].(\lni[2].\en[1].Z + \en[1].X))$}\\[2pt]
  \plat{$Z\hspace{-.35pt} \stackrel{{\it def}}{=} \lni[1].\ec[2].\lc[2].\en[2].Y + \ec[2].(\lni[1].\lc[2].\en[2].Y +
  \lc[2].(\lni[1].\en[2].Y + \en[2].X))$}\\
is  depicted on the right.
It satisfies ME\,1--5 with ${\it CC} \mathbin{:=} {\it Pr}$ and ME\,6 with ${\it CC} \mathbin{:=} WF(\Tsk)$,
where $\textsc{ln}_1,\textsc{ln}_2 \in \Tsk$.
It could be seen as a mediator that synchronises, on the actions $\lni$, $\ec$, $\lc$ and $\en$,
with the actual processes that need to exclusively enter their critical sections.
Yet, it would not be commonly accepted as a valid mutual exclusion protocol, since nothing prevents
it to never choose $\lni[2]$ when an alternative is available. This means that merely requiring weak
fairness in ME\,6 makes this requirement unacceptably weak.
The problem with this protocol is that it ensures \emph{Liveness} by making it hard for processes
to leave their noncritical sections.
\par}

\section{Reactive CTL}\label{sec:CTL}

This section presents a reactive version of \emph{Computation Tree Logic} (CTL) \cite{EC82}.
This shows that the ideas presented here are not specific to a linear-time logic.
The syntax of CTL is
\[\phi,\psi ::= p \mid \neg \phi \mid \phi \wedge \psi  \mid {\bf EX}\phi \mid {\bf AX}\phi \mid 
                {\bf EF}\phi \mid {\bf AF}\phi \mid {\bf EG}\phi \mid  {\bf AG}\phi \mid
                {\bf E} \psi {\bf U} \phi \mid {\bf A} \psi {\bf U} \phi \]
with $p \mathbin\in AP$ an atomic predicate.
The relation $\models$ between states $s$ in a Kripke structure, CTL formulae $\phi$ and completeness
criteria $CC$ is inductively defined by
\begin{itemize}
\item $s \models^{CC} p$, with $p \in AP$, iff $(s,p) \in {\models}$,
\item $s \models^{CC} \neg\phi$ iff $s \not\models^{CC} \phi$, 
\item $s \models^{CC} \phi \wedge \psi$ iff $s \models^{CC} \phi$ and $s \models^{CC} \psi$, 
\item $s \models^{CC} {\bf EX}\phi$ iff there is a state $s'$ with $s \rightarrow s'$ and $s' \models^{CC} \phi$,
\item $s \models^{CC} {\bf AX}\phi$ iff for each state $s'$ with $s \rightarrow s'$ one has $s' \models^{CC} \phi$,
\item $s \models^{CC} {\bf EF}\phi$ iff some complete path starting in $s$ contains a state $s'$ with $s' \models^{CC} \phi$,
\item $s \models^{CC} {\bf AF}\phi$ iff each complete path starting in $s$ contains a state $s'$ with $s' \models^{CC} \phi$,
\item $s \models^{CC} {\bf EG}\phi$ iff all states $s'$ on some complete path starting in $s$ satisfy $s' \models^{CC} \phi$,
\item $s \models^{CC} {\bf AG}\phi$ iff all states $s'$ on all complete paths starting in $s$ satisfy $s' \models^{CC} \phi$,
\item $s \models^{CC} {\bf E}\psi {\bf U}\phi$ iff some complete path $\pi$ starting in $s$ contains a state
  $s'$ with $s' \models^{CC} \phi$, and each state $s''$ on $\pi$ prior to $s'$ satisfies $s'' \models^{CC} \psi$,
\item $s \models^{CC} {\bf A}\psi {\bf U}\phi$ iff each complete path $\pi$ starting in $s$ contains a state
  $s'$ with $s' \models^{CC} \phi$, and each state $s''$ on $\pi$ prior to $s'$ satisfies $s'' \models^{CC} \psi$.
\end{itemize}
Exactly as for {\LTLX}, this allows the formulation of CTL judgements $s \models^{CC}_B \phi$.

\section{Conclusion}

I proposed a formalism for making temporal judgements $P \models^{CC}_B \phi$, with $P$ a process
specified in any formalism that admits a translation into LTSs, $\phi$ a temporal formula from a
logic like LTL or CTL, $CC$ a completeness criterion, stating which paths in the LTS model complete
system runs, and $B$ the set of actions that may be blocked by the user or environment of a
system. I applied this formalism to unambiguously express the requirements defining fair schedulers
and mutual exclusion protocols.

\bibliographystyle{eptcs}
\bibliography{../../../Stanford/lib/abbreviations,../../../Stanford/lib/new,../../../Stanford/lib/dbase,glabbeek}
\end{document}